\documentclass[prd,twocolumn,groupedaddress,showpacs,nofootinbib]{revtex4}
\usepackage{graphicx}
\usepackage{dcolumn}
\usepackage{bm}
\usepackage{amssymb}
\usepackage{mathrsfs}
\usepackage{amsmath}
\usepackage{epsfig}
\usepackage[dvips]{color}
\usepackage{hhline}

\begin{document}

\title{From dark matter to neutrinoless double beta decay}

\author{Pei-Hong Gu}
\email{peihong.gu@mpi-hd.mpg.de}

\affiliation{Max-Planck-Institut f\"{u}r Kernphysik, Saupfercheckweg
1, 69117 Heidelberg, Germany}

\begin{abstract}

Associated with two TeV-scale leptoquark scalars, a dark matter
fermion which is the neutral component of an isotriplet can mediate
a testable neutrinoless double beta decay at one-loop level. The
dark matter fermion with determined mass and spin-independent
scattering can be verified by the future dark matter direct
detection experiments. We also discuss the implications on neutrino
masses and baryon asymmetry.

\end{abstract}


\maketitle

\section{Introduction}

A process of neutrinoless double beta decay requires a lepton number
violation of two units. Such lepton number violation can have
various origins \cite{rodejohann2011,bm2002,gu2011}. For example,
the standard picture of the neutrinoless double beta decay is
realized by the electron neutrino with a tiny Majorana mass. The
smallness of neutrino masses can be understood by the seesaw
mechanism \cite{minkowski1977,mw1980,flhj1989,barr2003} or some loop
and chirality suppression factors
\cite{zee1980,zee1985,mahanta1999,knt2002}. Neutrino physics may be
related to other interesting topics in particle physics and
cosmology such as dark matter \cite{knt2002,ma2006} and inflation
\cite{adm2007}. In the case that the neutrino masses arise from
certain loop diagrams involving the dark matter particle
\cite{knt2002,ma2006} and then result in a standard neutrinoless
double beta decay, we can say the neutrinoless double beta decay is
indirectly related to the dark matter.

In this paper we shall propose a novel scenario where the dark
matter particle is directly responsible for generating the
neutrinoless double beta decay. Specifically, we shall extend the
$SU(3)_{c}^{}\times SU(2)_{L}^{}\times U(1)_{Y}^{}$ standard model
(SM) by one $[SU(2)^{}_L]$-triplet fermion and two
$[SU(2)^{}_L]$-doublet leptoquark scalars. There is a $Z^{}_2$
discrete symmetry under which the fermion triplet and one of the
leptoquarks are odd while the other fields are even. The fermion
triplet without nonzero hypercharges has a Majorana mass to break
the lepton number by two units. The $Z^{}_2$-odd leptoquark has the
Yukawa couplings to the fermion triplet and the quark doublets while
the $Z^{}_2$-even leptoquark has the Yukawa couplings to the lepton
doublets and the down-type quark singlets. The neutral component of
the fermion triplet should have a determined mass around
$2.4\,\textrm{TeV}$ if it is the dark matter particle. Associated
with the leptoquarks, the dark matter fermion can mediate a
neutrinoless double beta decay at one-loop level. Meanwhile, the
neutrino masses can be induced at three- and five-loop level. By
taking the leptoquarks at the TeV scale, the neutrinoless double
beta decay can arrive at a testable level. The neutrinoless double
beta decay and other lepton number violating processes will go out
of equilibrium at a temperature about $100\,\textrm{GeV}$, below
which the cosmological baryon asymmetry is allowed to produce. As
for the loop-induced neutrino masses, their magnitude will not be
bigger than the eV order. The dark-matter-nucleon scattering can be
verified by the future experiments since it has a determined
spin-independent cross section about
$10^{-45}_{}\,\textrm{cm}^2_{}$, besides a spin-dependent cross
section smaller than $\mathcal{O}(10^{-46}_{}\,\textrm{cm}^2_{})$.

\section{The model}

For simplicity, we will not write down the full Lagrangian. Instead,
we only give the following terms relevant to our demonstration,
\begin{eqnarray}
\label{lagrangian}\mathcal{L}&\supset&
-h_{i}^{}\bar{q}_{L_i^{}}^{c}i\tau_2^{}T_{L}^{}\eta-f_{ij}^{}\bar{l}_{L_i^{}}^{}\xi
d_{R_j^{}}^{}-\lambda(\xi^\dagger_{}\eta)^2_{}+\textrm{H.c.}\nonumber\\
&&
-\frac{1}{2}M_{T}^{0}[\textrm{Tr}(\bar{T}_{L}^{c}T_{L}^{})+\textrm{H.c.}]\,,
\end{eqnarray}
where we have defined the $[SU(2)]$-triplet fermion:
\begin{eqnarray}
T_{L}^{}(\textbf{1},\textbf{3},0)&=&
\left[\begin{array}{rr}\frac{1}{\sqrt{2}}T_{L}^{0}&T_{L}^{+}\\
[2mm]T_{L}^{-} & -\frac{1}{\sqrt{2}}T_{L}^{0}
\end{array}\right]\,,
\end{eqnarray}
and the $[SU(2)]$-doublet leptoquark scalars:
\begin{eqnarray}
\eta^{}(\bar{\textbf{3}},\textbf{2},-\frac{1}{6})=\left[\begin{array}{r}\eta^{+\frac{1}{3}}_{}\\
[1mm] \eta^{-\frac{2}{3}}_{}\end{array}\right]\,,
~~\xi^{}(\bar{\textbf{3}},\textbf{2},-\frac{1}{6})=\left[\begin{array}{r}\xi^{+\frac{1}{3}}_{}\\
[1mm] \xi^{-\frac{2}{3}}_{}\end{array}\right]\,,
\end{eqnarray}
besides the SM quarks and leptons:
\begin{eqnarray}
&&q_{L_i^{}}^{}(\textbf{3},\textbf{2},+\frac{1}{6})=\left[\begin{array}{l}u_{L_i^{}}^{}\\
[2mm] d_{L_i^{}}^{}\end{array}\right]\,,~~d_{R_i^{}}^{}(\textbf{3},\textbf{1},-\frac{1}{3})\,,\nonumber\\
&&l_{L_i^{}}^{}(\textbf{1},\textbf{2},-\frac{1}{2})=\left[\begin{array}{l}\nu_{L_i^{}}^{}\\
[2mm] e_{L_i^{}}^{}\end{array}\right]~~\textrm{with}~~i=1,2,3\,.
\end{eqnarray}
It should be noted that our model respects a $Z_2^{}$ discrete
symmetry under which only the fermion triplet $T_L^{}$ and the
leptoquark $\eta$ carry an odd parity.

\section{Dark matter fermion}

Although the charged and neutral components of the fermion triplet
$T_L^{}$ have a same mass at tree level, i.e.
\begin{eqnarray}
\mathcal{L}&\supset& -M_{T}^0
\overline{T^{-}_{}}T^{-}_{}-\frac{1}{2}M_{T}^0
\overline{T^0_{}}T^0_{}~~\textrm{with}\nonumber\\
&&T^{\pm}_{}=T_{L}^{\pm}+(T_{L}^{\mp})^c_{}\,,~~T^0_{}=T_{L}^{0}+(T_{L}^{0})^c_{}\,,
\end{eqnarray}
the one-loop electroweak corrections will make the charged
$T^{\pm}_{}$ to be heavier than the neutral $T^0_{}$ \cite{cfs2005},
\begin{eqnarray}
M_{T^{\pm}_{}}^{}-M_{T^0_{}}^{}\simeq
\frac{g^2_{}}{4\pi}m_W^{}\sin^2_{}\frac{\theta_W^{}}{2}=166\,\textrm{MeV}\,.
\end{eqnarray}
So, the neutral $T^{0}_{}$ can keep stable and leave a relic density
to the present universe, if it is lighter than the leptoquark
$\eta$.

The relic density is determined by the annihilations of the neutral
$T^{0}_{}$ and the charged $T^{\pm}_{}$ \cite{ms2008,cfs2005}. The
effective cross section contains two parts,
\begin{eqnarray}
\langle\sigma_A^{}v_{\textrm{rel}}^{}\rangle=\langle\sigma_A^{}v_{\textrm{rel}}^{}\rangle_{g}^{}
+\langle\sigma_A^{}v_{\textrm{rel}}^{}\rangle_{Y}^{}\,,
\end{eqnarray}
where the first term is from the gauge interactions
\cite{ms2008,cfs2005},
\begin{eqnarray}
\label{gcs}
\langle\sigma_A^{}v_{\textrm{rel}}^{}\rangle_{g}^{}\simeq
\frac{37g^4_{}}{192\pi}\frac{1}{M_{T^0_{}}^2}\,.
\end{eqnarray}
while the second term is from the Yukawa interactions,
\begin{eqnarray}
\label{ycs}
\langle\sigma_A^{}v_{\textrm{rel}}^{}\rangle_{Y}^{}\simeq
\frac{(h^\dagger_{}h)^2_{}}{12\pi}
\frac{M_{T^0_{}}^2}{[M_{T^0_{}}^2+M_{\eta}^2]^2_{}}\,.
\end{eqnarray}
Here $v_{\textrm{rel}}^{}$ is the relative velocity between the two
annihilating fermions in their cms system, while $M_\eta^{}$ is the
mass of the leptoquark $\eta$. If the neutral $T^0_{}$ accounts for
the dark matter relic density \cite{ma2005}, it should have a
determined mass about $M_{T^0_{}}^{}= 2.4\,\textrm{TeV}$
\cite{ms2008,cfs2005} in the absence of the Yukawa contribution
(\ref{ycs}). The dark matter mass will increase if the Yukawa
contribution is taken into account. Actually, we find
\begin{eqnarray}
\frac{\langle\sigma_A^{}v_{\textrm{rel}}^{}\rangle_{Y}^{}}{\langle\sigma_A^{}v_{\textrm{rel}}^{}\rangle_{g}^{}}
&\simeq&
\frac{16(h^\dagger_{}h)^2_{}}{37g^4_{}}\frac{M_{T^0_{}}^4}{[M_{T^0_{}}^2+M_{\eta}^2]^2_{}}\ll
1~~\textrm{for}\nonumber\\
&&h_{i}^{}\lesssim \mathcal{O}(1)\,,~~M_{T^0_{}}^{4}\ll
M_{\eta}^{4}\,.
\end{eqnarray}
So, the dark matter mass can be fixed by
\begin{eqnarray}
M_{T^0_{}}^{}= 2.4\,\textrm{TeV}\,.
\end{eqnarray}

At one-loop level, the scattering of the dark matter fermion
$T^0_{}$ on the nucleons $N$ can have a spin-independent cross
section \cite{cfs2005}:
\begin{eqnarray}
\sigma_{\textrm{SI}}^{N}
&=&\frac{g^8_{}}{256\pi^3_{}}\frac{f^2_{N}m_N^4}{m_W^6}\left(1+\frac{m_W^2}{m_H^2}\right)^2_{}~~\textrm{with}\nonumber\\
&&f_N^{}=\sum_{q=u,d,s}^{}f^{(N)}_{Tq}+\frac{2}{27}\sum_{q=c,b,t}^{}f^{(N)}_{Tq}\,.
\end{eqnarray}
By inputting \cite{nakamura2010}
\begin{eqnarray}
&&g^4_{}=32 G_F^2
m_{W}^4=0.182\,,~~m_W^{}=80.398\,\textrm{GeV}\,,\nonumber\\
&&m_p^{}=938.3\,\textrm{MeV}\,,~~m_n^{}=939.6\,\textrm{MeV}\,,
\end{eqnarray}
and \cite{efo2000}
\begin{eqnarray}
&&f_{Tu}^{(p)}=0.020\,,~f_{Td}^{(p)}=0.026\,,~f_{Ts}^{(p)}=0.118\,,\nonumber\\
&&f_{Tc}^{(p)}=f_{Tb}^{(p)}=f_{Tt}^{(p)}=1-\sum_{q=u,d,s}^{}f^{(p)}_{Tq}=0.836\,,\nonumber\\
&&f_{Tu}^{(n)}=0.014\,,~f_{Td}^{(n)}=0.036\,,~f_{Ts}^{(n)}=0.118\,,\nonumber\\
&&f_{Tc}^{(n)}=f_{Tb}^{(n)}=f_{Tt}^{(n)}=1-\sum_{q=u,d,s}^{}f^{(n)}_{Tq}=0.832\,,
\end{eqnarray}
we can read
\begin{eqnarray}
\sigma_{\textrm{SI}}^{N}&\simeq&
10^{-45}_{}\,\textrm{cm}^2_{}~~\textrm{for}~~
m_H^{}=126\,\textrm{GeV}\,.
\end{eqnarray}
So, the spin-independent scattering can be verified by the future
dark matter direct detection experiments such as the superCDMS
experiment \cite{schnee2005}.

We can also have a spin-dependent dark-matter-nucleon scattering
through the tree-level exchange of the leptoquark $\eta$. The
relevant axial-vector interactions are given by
\begin{eqnarray}
\mathcal{L}&\supset& -\frac{h^{}_{j}h^{\ast}_{i}}{16M_{\eta}^2}
(\overline{T^0_{}}\gamma^\mu_{}\gamma^5_{}T^0_{})(\bar{d}_i^{}\gamma_\mu^{}\gamma^5_{}d_j^{}
+\bar{u}_i^{}\gamma_\mu^{}\gamma^5_{}u_j^{})\nonumber\\
&&+\textrm{H.c.}\,,
\end{eqnarray}
from which we can write the effective dark-matter-nucleon
interaction to be
\begin{eqnarray}
\mathcal{L}&\supset& -a^{}_N (\overline{T^0_{}}
\gamma^\mu_{}\gamma^5_{}T^0_{}) (\bar{N} s^{(N)}_\mu
N)~~\textrm{with}\nonumber\\
&&a_N^{}=\frac{1}{8
M_\eta^2}\{|h_{1}^{}|^2[\Delta_u^{(N)}+\Delta_d^{(N)}]+|h_{2}^{}|^2\Delta_s^{(N)}\}\,.\nonumber\\
&&
\end{eqnarray}
Here $s^{(N)}_\mu$ is the spin of the nucleon $N$, and
$\Delta_q^{(N)}$ are extracted from the data on polarized deep
elastic scattering. The spin-dependent cross section quoted by the
experiments can be calculated by \cite{cpp2010}
\begin{eqnarray}
\sigma_{\textrm{SD}}^{N}
&=&\frac{6}{\pi}\mu^2_{r}a_N^2~~\textrm{with}~~\mu_r^{}=\frac{M_{T^0_{}}^{}m_N^{}}{M_{T^0_{}}^{}+m_N^{}}\,.
\end{eqnarray}
By taking \cite{efo2000}
\begin{eqnarray}
&&\Delta_u^{(p)}=\Delta_d^{(n)}=0.78\,,~\Delta_d^{(p)}=\Delta_u^{(n)}=-0.48\,,\nonumber\\
&&\Delta_s^{(p)}=\Delta_s^{(n)}=-0.15\,,
\end{eqnarray}
and assuming
\begin{eqnarray}
M_\eta^{}=3\,M_{T^0_{}}^{}\,,
\end{eqnarray}
we can obtain
\begin{eqnarray}
\sigma_{\textrm{SD}}^{N}
&=&\left(\frac{0.3|h_{1}^{}|^2_{}-0.15|h_{2}^{}|^2_{}}{0.15}\right)^2_{}\times
10^{-46}_{}\,\textrm{cm}^2_{}\nonumber\\
&\lesssim&\mathcal{O}(10^{-46}_{}\,\textrm{cm}^2_{})~~\textrm{for}~~h_{1}^{},h_{2}^{}\lesssim
\mathcal{O}(1)\,,
\end{eqnarray}
which is far below the experimental limits \cite{ksz2010}.

\section{Neutrinoless double beta decay}

\begin{figure}
\vspace{6.5cm} \epsfig{file=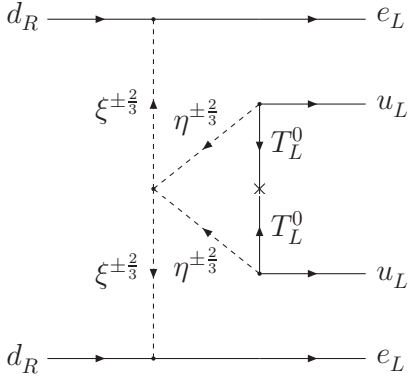, bbllx=5.3cm, bblly=4.0cm,
bburx=15.3cm, bbury=14cm, width=8cm, height=8cm, angle=0, clip=0}
\vspace{-9cm} \caption{\label{nuless} One-loop neutrinoless double
beta decay.}
\end{figure}

As show in Fig. \ref{nuless}, the dark matter fermion $T^0_{}$
associated with the leptoquark scalars $\eta^{\pm\frac{2}{3}}_{}$
and $\xi^{\pm\frac{2}{3}}_{}$ can mediate a one-loop diagram to
generate a neutrinoless double beta decay. The effective operators
are given by
\begin{eqnarray}
\label{nuless} \mathcal{L}&\supset&-G^{}_{0\nu}
(\overline{d_R^c}d_R^{})(\overline{u_L^{}}e_L^{c})(\overline{u_L^{}}e_L^{c})\nonumber\\
&=&-\frac{1}{4}G_{0\nu}^{}(\overline{e_L^{}}e^c_{L})[(\overline{u_L^{}}d_R^{})
(\overline{u_L^{}}d_R^{})
\nonumber\\
&&-(\overline{u_L^{}}\sigma_{\mu\nu}^{}d_R^{})
(\overline{u_L^{}}\sigma^{\mu\nu}_{}d_R^{})]+\textrm{H.c.} \,,
\end{eqnarray}
with the coefficients:
\begin{eqnarray}
G_{0\nu}^{}&=&i\frac{\lambda
(h_{1}^\ast)^2_{}f^2_{11}}{4\pi^2_{}}\frac{M_{T^0_{}}^{}}{M_{\xi}^4(M_{\eta}^2-M_{T^0_{}}^2)}\nonumber\\
&& \times
\left(1-\frac{M_{T^0_{}}^2}{M_{\eta}^2-M_{T^0_{}}^2}\ln\frac{M_{\eta}^2}{M_{T^0_{}}^2}\right)\,.
\end{eqnarray}
Here $M_\xi^{}$ is the mass of the leptoquark $\xi$. We need
calculate the half-life of the neutrinoless double beta decay to
compare with the experimental limits. For this purpose, we rewrite
the low energy Lagrangian (\ref{nuless}) to be
\begin{eqnarray}
\mathcal{L}^{}_{}&\supset&\frac{G_F^{2}}{2m_p^{}}[\bar{e}(1+\gamma_5^{})e^c_{}]\{\eta_{}^{PS}[\bar{u}(1+\gamma_5^{})d]
[\bar{u}(1+\gamma_5^{})d]
\nonumber\\
&&-\frac{1}{4}\eta_{}^{T}[\bar{u}\sigma_{\mu\nu}^{}(1+\gamma_5^{})d]
[\bar{u}\sigma^{\mu\nu}_{}(1+\gamma_5^{})d]\}+\textrm{H.c.}\,,\quad
\end{eqnarray}
where the effective lepton number violating parameters
$\eta_{}^{PS}$ and $\eta_{}^{T}$ are defined by
\begin{eqnarray}
\eta_{}^{PS}=\frac{m_p^{}G_{0\nu}^{}}{16G_F^{2}}\,,
~~\eta_{}^{T}=\frac{m_p^{}G_{0\nu}^{}}{4G_F^{2}}\,.
\end{eqnarray}

The experimental lower bound $T_{1/2}^{exp}(Y)$ on the half-life of
a certain isotope $Y$ can constrain the effective lepton number
violating parameters \cite{fks1998},
\begin{eqnarray}
\eta=\frac{5}{8}\eta_{}^{PS}+\frac{3}{8}\eta_{}^{T}\leq
\eta_{\textrm{X}}^{exp}=\frac{10^{-7}_{}}{\zeta(Y)}\sqrt{\frac{10^{24}_{}\,\textrm{yrs}}{T_{1/2}^{exp}(Y)}}\,,
\end{eqnarray}
where the quantity $\zeta(Y)$ is an intrinsic characteristic of the
isotope $Y$. By taking $G_F^{}=1.16637\times
10^{-5}_{}\,\textrm{GeV}^{-2}_{}$ \cite{nakamura2010},
$\zeta(^{76}_{}\textrm{Ge})=5.5$ \cite{fks1998} and
$T_{1/2}^{exp}(^{76}_{}\textrm{Ge})=1.9\times
10^{19}_{}\,\textrm{yrs}$ \cite{rodejohann2011}, we can find
\begin{eqnarray}
G_{0\nu}^{}\leq 3.1\times 10^{-18}_{}\,\textrm{GeV}^{-5}_{}\,.
\end{eqnarray}
To fulfill the above constraint, we can consider a proper parameter
choice such as
\begin{eqnarray}
G_{0\nu}^{}&=&i \left(\frac{\lambda}{1}\right)\left(\frac{
h^{\ast}_{1}}{1}\right)^2_{}\left(\frac{f^{}_{11}}{1}\right)^2_{}
\times
\left(\frac{750\,\textrm{GeV}}{M_{\xi}^{}}\right)^4_{}\nonumber\\
&&\times3.0\times
10^{-18}_{}\,\textrm{GeV}^{-5}_{}~\textrm{for}~M_\eta^{}=3M_{T^0_{}}^{}\,.
\end{eqnarray}

\section{Implications on neutrino masses and baryon asymmetry}

\begin{figure*}
\vspace{5.0cm} \epsfig{file=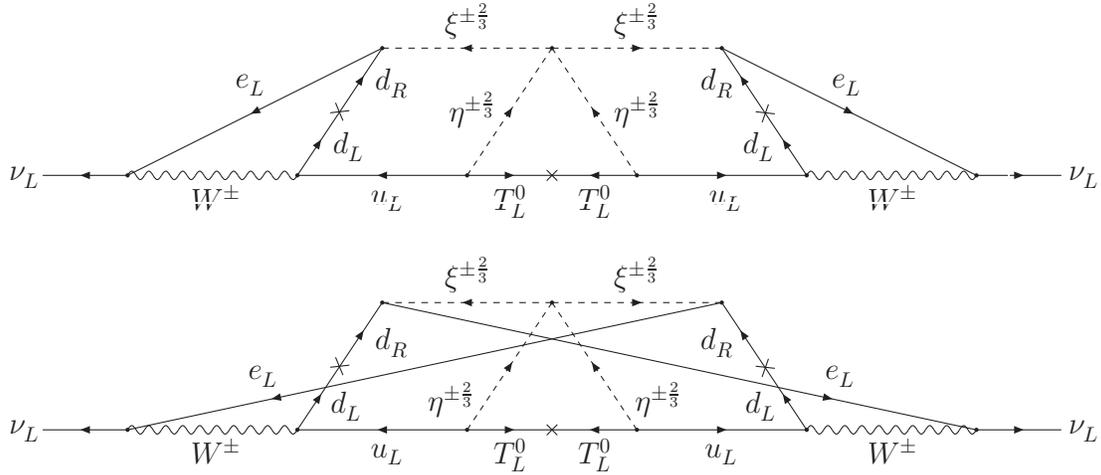, bbllx=5.6cm, bblly=6.0cm,
bburx=15.6cm, bbury=16cm, width=8cm, height=8cm, angle=0, clip=0}
\vspace{-6.5cm} \caption{\label{numass5} Five-loop neutrino mass
generation.}
\end{figure*}

\begin{figure*}
\vspace{1.0cm} \epsfig{file=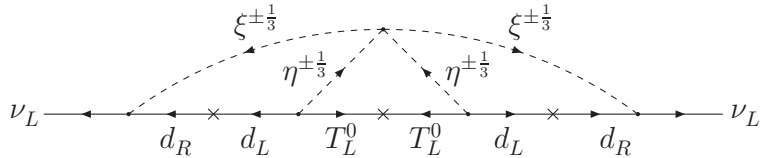, bbllx=5.6cm, bblly=6.0cm,
bburx=15.6cm, bbury=16cm, width=8cm, height=8cm, angle=0, clip=0}
\vspace{-6.5cm} \caption{\label{numass3} Three-loop neutrino mass
generation.}
\end{figure*}

Any neutrinoless double beta decay processes will eventually result
in a Majorana neutrino mass term according to the Schechter-Valle
theorem \cite{sv1982}. For the present one-loop neutrinoless double
beta decay, the leading neutrino masses should appear at five-loop
level as shown in Fig. \ref{numass5}. If the interactions involving
the second- and third-generation charged fermions are also taken
into account, we can obtain a $3\times 3$ neutrino mass matrix as
below,
\begin{eqnarray}
m_{\nu}^{5-\textrm{loop}}\sim \frac{4g^4_{}}{(16\pi^2)^5_{}}\lambda
(f \hat{m}_d^{} V^\dagger_{} h^\ast)\frac{M^{}_{T^0_{}}}{M_\eta^2}(f
\hat{m}_d^{} V^\dagger_{} h^\ast)^T_{}\,,
\end{eqnarray}
with $\hat{m}_d^{}=\textrm{diag}\{m_d^{}\,,m_s^{}\,,m_b^{}\}$ being
the down-type quark masses and $V$ being the
Cabibbo-Kobayashi-Maskawa \cite{cabibbo1963} (CKM) matrix. Our model
can also generate the Majorana neutrino masses at three-loop level
even if we don't resort to the gauge interactions. The relevant
diagram is shown in Fig. \ref{numass3}. We can estimate the
three-loop neutrino masses to be
\begin{eqnarray}
m_{\nu}^{3-\textrm{loop}}&\sim& \frac{4}{(16\pi^2)^3_{}}\lambda(f
\hat{m}_d^{} h^\ast)\frac{M^{}_{T^0_{}}}{M_\eta^2}(f \hat{m}_d^{}
h^\ast)^T_{}\,.
\end{eqnarray}

It is easy to check that the neutrino masses dominated by the
three-loop contribution can arrive at an acceptable level, i.e.
\begin{eqnarray}
(m_{\nu}^{3-\textrm{loop}})_{ij}^{}&\sim&
\left(\frac{\lambda}{1}\right)\left(\frac{
h^{\ast}_{3}}{1}\right)^2_{}\left(\frac{f^{}_{i3}}{1}\right)\left(\frac{f^{}_{j3}}{1}\right)
\left(\frac{m_{b}^{}}{4.2\,\textrm{GeV}}\right)^2_{}\nonumber\\
&&\times 0.8\,\textrm{eV}~~\textrm{for}~~M_\eta^{}=3M_{T^0_{}}^{}\,.
\end{eqnarray}
We should keep in mind that the neutrino mass matrix is rank-1 and
only has one nonzero eigenvalue so that it cannot explain the
neutrino oscillation data. In other words, the neutrino masses
should have additional sources such as the conventional seesaw.

The neutrinoless double beta decay and other lepton number violating
processes will wash out any lepton asymmetries until they go out of
equilibrium at the temperature estimated by
\begin{eqnarray}
&&\Gamma\sim G_{0\nu}^2 T^{11}=H(T) \Rightarrow \nonumber\\
&&\quad T=
100\,\textrm{GeV}\left(\frac{10^{-18}_{}\,\textrm{GeV}^{-5}_{}}{G_{0\nu}^{}}\right)^{\frac{2}{9}}_{}
\left[\frac{g_{\ast}^{}(T)}{100}\right]^{\frac{1}{18}}_{}\,.\quad
\end{eqnarray}
Here
\begin{eqnarray}
H(T)=\left[\frac{8\pi^{3}_{}g_{\ast}^{}(T)}{90}\right]^{\frac{1}{2}}_{}
\frac{T^{2}_{}}{M_{\textrm{Pl}}^{}}
\end{eqnarray}
is the Hubble constant with $M_{\textrm{Pl}}^{}\simeq 1.22\times
10^{19}_{}\,\textrm{GeV}$ being the Planck mass and $g_\ast^{}(T)$
being the relativistic degrees of freedom. These lepton number
violating processes will also wash out the existing baryon asymmetry
in the presence of the $SU(2)_L^{}$ sphaleron processes
\cite{krs1985}. The cosmological baryon asymmetry thus should be
produced below the temperature $T\simeq 100\,\textrm{GeV}$.

\section{Conclusion and discussion}

In this paper we have shown that the dark matter particle can be
directly responsible for generating the neutrinoless double beta
decay. In our model, the dark matter fermion associated with two
TeV-scale leptoquark scalars can simultaneously mediate the
neutrinoless double beta decay at one-loop level and the neutrino
masses at three- and five-loop level. The neutrinoless double beta
decay can arrive at a testable level, while the accompanying
neutrino masses will not exceed the eV order. The neutrinoless
double beta decay and other lepton number violating processes will
constrain the cosmological baryon asymmetry to produce at a scale
below $100\,\textrm{GeV}$. The dark matter fermion with the
determined mass and spin-independent scattering cross section can be
verified by the future dark matter direct detection experiments.

The present model can be modified. For example, we can replace the
$[SU(2)^{}_L]$-doublet leptoquarks
$\eta(\bar{\textbf{3}},\textbf{2},-\frac{1}{6},-)$ and
$\xi(\bar{\textbf{3}},\textbf{2},-\frac{1}{6},+)$ by two
$SU(2)^{}_L$ triplets
$\Delta(\bar{\textbf{3}},\textbf{3},+\frac{1}{3},-)$ and
$\Omega(\bar{\textbf{3}},\textbf{3},+\frac{1}{3},+)$, or by one
$SU(2)^{}_L$ triplet
$\Delta(\bar{\textbf{3}},\textbf{3},+\frac{1}{3},-)$ and one
$SU(2)^{}_L$ singlet
$\omega(\bar{\textbf{3}},\textbf{1},+\frac{1}{3},+)$. Here and
thereafter "$+$" and "$-$" denote the parity under the $Z^{}_2$
discrete symmetry. Alternatively, the dark matter fermion may be a
gauge-singlet fermion $S^{}_R(\textbf{1},\textbf{1},0,-)$. In this
case, the leptoquarks should be (i) two $SU(2)^{}_L$ doublets
$\eta(\bar{\textbf{3}},\textbf{2},-\frac{1}{6},-)$ and
$\xi(\bar{\textbf{3}},\textbf{2},-\frac{1}{6},+)$, (ii) two
$SU(2)^{}_L$ singlets
$\delta(\bar{\textbf{3}},\textbf{1},+\frac{1}{3},-)$ and
$\omega(\bar{\textbf{3}},\textbf{1},+\frac{1}{3},+)$, (iii) one
$SU(2)^{}_L$ singlet
$\delta(\bar{\textbf{3}},\textbf{1},+\frac{1}{3},-)$ and one
$SU(2)^{}_L$ triplet
$\Omega(\bar{\textbf{3}},\textbf{3},+\frac{1}{3},+)$.

\textbf{Acknowledgement}: This work is supported by the
Sonderforschungsbereich TR 27 of the Deutsche
Forschungsgemeinschaft.


\begin{thebibliography}{99}




\bibitem{rodejohann2011}
For recent reviews, see W. Rodejohann, Int. J. Mod. Phys. E
\textbf{20}, 1833 (2011); J.D. Vegados, H. Ejiri, and F.
\u{S}imkovic, arXiv:1205.0649 [hep-ph].

\bibitem{bm2002}
B. Brahmachari and E. Ma, Phys. Lett. B \textbf{536}, 259 (2002).

\bibitem{gu2011}
P.H. Gu, Phys. Rev. D \textbf{85}, 093016 (2012).

\bibitem{minkowski1977}
P. Minkowski, Phys. Lett. B \textbf{67}, 421 (1977); T. Yanagida, in
{\it Proceedings of the Workshop on Unified Theory and the Baryon
Number of the Universe}, edited by O. Sawada and A. Sugamoto (KEK,
Tsukuba, 1979), p. 95; M. Gell-Mann, P. Ramond, and R. Slansky, in
{\it Supergravity}, edited by F. van Nieuwenhuizen and D. Freedman
(North Holland, Amsterdam, 1979), p. 315; S.L. Glashow, in {\it
Quarks and Leptons}, edited by M. L\'{e}vy {\it et al.} (Plenum, New
York, 1980), p. 707; R.N. Mohapatra and G. Senjanovi\'{c}, Phys.
Rev. Lett. \textbf{44}, 912 (1980).


\bibitem{mw1980}
M. Magg and C. Wetterich, Phys. Lett. B \textbf{94}, 61 (1980); J.
Schechter and J.W.F. Valle, Phys. Rev. D \textbf{22}, 2227 (1980);
T.P. Cheng and L.F. Li, Phys. Rev. D \textbf{22}, 2860 (1980); G.
Lazarides, Q. Shafi, and C. Wetterich, Nucl. Phys. B \textbf{181},
287 (1981); R.N. Mohapatra and G. Senjanovi\'{c}, Phys. Rev. D
\textbf{23}, 165 (1981).


\bibitem{flhj1989}
R. Foot, H. Lew, X.G. He, and G.C. Joshi, Z. Phys. C \textbf{44},
441 (1989).


\bibitem{barr2003}
S.M. Barr, Phys. Rev. Lett. \textbf{92}, 101601 (2004).





\bibitem{zee1980}
A. Zee, Phys. Lett. B \textbf{93}, 389 (1980).


\bibitem{zee1985}
A. Zee, Phys. Lett. B \textbf{161}, 141 (1985); K.S. Babu, Phys.
Lett. B \textbf{203}, 132 (1988).

\bibitem{mahanta1999}
U. Mahanta, Phys. Rev. D \textbf{62}, 073009 (2000); K.S. Babu and
C.N. Leung, Nucl. Phys. B \textbf{619}, 667 (2001); A. de Gouv\^{e}a
and J. Jenkins, Phys. Rev. D \textbf{77}, 013008 (2008); D.
Aristizabal Sierra, M. Hirsch, and S.G. Kovalenko, Phys. Rev. D
\textrm{77}, 055011 (2008); K.S. Babu and J. Julio, Nucl. Phys. B
\textbf{841}, 130 (2010); K.S. Babu and J. Julio, Phys. Rev. D
\textbf{85}, 073005 (2012).

\bibitem{knt2002}
L.M. Krauss, S. Nasri, and M. Trodden, Phys. Rev. D \textbf{67},
085002 (2003); K. Cheung and O. Seto, Phys. Rev. D \textbf{69},
113009 (2004).

\bibitem{ma2006}
E. Ma, Phys. Rev. D \textbf{73}, 077301 (2006).


\bibitem{adm2007}
R. Allahverdi, B. Dutta, and A. Mazumdar, Phys. Rev. Lett.
\textbf{99}, 261301 (2007); A. Mazumdar and S. Morisi, Phys. Rev. D
\textbf{86}, 045031 (2012).


\bibitem{cfs2005}
M. Cirelli, N. Fornengo, and A. Strumia, Nucl. Phys. B \textbf{753},
178 (2006).


\bibitem{ms2008}
E. Ma and D. Suematsu, Mod. Phys. Lett. A \textbf{24}, 583 (2009);
W. Cao, arXiv:1202.6394 [hep-ph].


\bibitem{ma2005}
E. Ma, Phys. Lett. B \textbf{625}, 76 (2005).






\bibitem{nakamura2010}
K. Nakamura {\it et al.}, (Particle Data Group), J. Phys. G
\textbf{37}, 075021 (2010).



\bibitem{efo2000}
J.R. Ellis, A. Ferstl, and K.A. Olive, Phys. Lett. B \textbf{481},
304 (2000).


\bibitem{schnee2005}
R.W. Schnee {\it et al.}, (superCDMS Collaboration),
astro-ph/0502435.


\bibitem{cpp2010}
T. Cohen, D.J. Phalen, and A. Pierce, Phys. Rev. D \textbf{81},
116001 (2010).



\bibitem{ksz2010}
J. Kopp, T. Schwetz, and J. Zupan, JCAP \textbf{1002}, 014 (2010).


\bibitem{fks1998}
A. Faessler, S. Kovalenko, and F. \u{S}imkovic, Phys. Rev. D
\textbf{58}, 115004 (1998).



\bibitem{sv1982}
J. Schechter and J.W.F. Valle, Phys. Rev. D \textbf{25}, 2951
(1982); J.F. Nieves, Phys. Lett. B \textbf{147}, 375 (1984); E.
Takasugi, Phys. Lett. B \textbf{149}, 372 (1984).



\bibitem{cabibbo1963}
N. Cabibbo, Phys. Rev. Lett. \textbf{10}, 531 (1963); M. Kobayashi
and T. Maskawa, Prog. Theor. Phys. \textbf{49}, 652 (1973).

\bibitem{krs1985}
V.A. Kuzmin, V.A. Rubakov, and M.E. Shaposhnikov, Phys. Lett. B
\textbf{155}, 36 (1985).

\end{thebibliography}
\end{document}